\documentclass[final,5p,times,twocolumn]{elsarticle}

\usepackage{amssymb}
\usepackage{mathtools}
\usepackage{tabularx}
\usepackage{siunitx}

\journal{Forces in Mechanics}

\begin{document}

\begin{frontmatter}



\title{LEFM\tnoteref{label5} is agnostic to geometrical nonlinearities arising at atomistic crack tips}

\author[label2]{Tarakeshwar Lakshmipathy\corref{cor1}}\ead{tara.ll.lakshmipathy@fau.de}
\author[label3]{Paul Steinmann}\ead{paul.steinmann@fau.de}
\author[label4]{Erik Bitzek}\ead{e.bitzek@mpie.de}
\cortext[cor1]{Corresponding author.}

\tnotetext[label5]{Linear Elastic Fracture Mechanics}

\fntext[label2]{Department of Materials Science and Engineering, Institute I, \& Competence Unit for Scientific Computing (CSC), Friedrich-Alexander-Universität Erlangen-Nürnberg (FAU), Martensstr. 5, 91058 Erlangen, Germany \ead{tara.ll.lakshmipathy@fau.de}}

\fntext[label3]{Department of Mechanical Engineering, Institute of Applied Mechanics, Friedrich-Alexander-Universität Erlangen-Nürnberg (FAU), Egerlandstraße 5, 91058 Erlangen, Germany \ead{paul.steinmann@fau.de}}

\fntext[label4]{Department Computational Materials Design, Max-Planck-Institut für Eisenforschung GmbH, Max-Planck-Straße 1, 40237 Düsseldorf, Germany \ead{e.bitzek@mpie.de}}

%
%




\begin{abstract}
Various fields such as mechanical engineering, materials science, etc., have seen a widespread use of linear elastic fracture mechanics (LEFM) at the continuum scale. 
LEFM is also routinely applied to the atomic scale. 
However, its applicability at this scale remains less well studied, with
most studies focusing on non-linear elastic effects.
Using a harmonic "snapping spring” nearest-neighbor potential which provides the closest match to LEFM on a discrete lattice, we show that the discrete nature of an atomic lattice leads to deviations from the LEFM displacement field during energy minimization. 
We propose that these deviations can be ascribed to geometrical nonlinearities since the material does not have a nonlinear elastic response prior to bond breaking. 
We demonstrate that crack advance and the critical stress intensity factor in an incremental loading scenario is governed by the collectively loaded region, and can not be determined analytically from the properties (max. elongation, max. sustained force, etc.) of the stressed crack tip bond alone.
\end{abstract}




\end{frontmatter}



\section{Introduction}
\label{intro}

Linear elastic fracture mechanics (LEFM) has a long history in structural integrity and the design of fracture-resistant materials, and is well established in literature \cite{RefAnderson}. 
The roots of LEFM trace back to the works of Inglis \cite{refInglis} who introduced the concept of a stress concentration factor to describe the stresses due to an elliptical hole with respect to an applied macroscopic stress. 
Later, Griffith \cite{RefGriffith} assumed a linear elastic material to establish a thermodynamic criterion for perfectly brittle crack advance. 
According to this model, a crack would propagate when the the stored elastic energy released by crack propagation exceeds the energy required to create two new crack surfaces. 
The energy release rate $G$, which can be defined as the rate of change in potential energy with crack area, can then be related to Griffith's criterion as follows:
\begin{equation}\label{form_griffith_err}
G \leq G_\mathrm{G} = 2\gamma ,
\end{equation}
where $G_\mathrm{G}$ is Griffith's theoretical resistance of the material that needs to be overcome to create two crack new surfaces, with $\gamma$ being their surface energy. Williams \cite{refWilliams} and Irwin \cite{RefIrwin} then used a stress-based approach to establish the concept of a stress intensity factor (SIF). This factor $K$ is a single loading parameter that describes the scaling of the amplitude of the stress field around the crack. The stress intensity factor is related to the energy release rate as follows:
\begin{equation}\label{form_general_sif}
K = \sqrt{GE^*} ,
\end{equation}
where $E^*$ is the orientation dependent elastic modulus. Applying the Griffith criterion to the stress-based approach using equations (\ref{form_griffith_err}) and (\ref{form_general_sif}), we obtain the theoretical SIF $K_\mathrm{G}$ required for crack advance:
\begin{equation}\label{form_griffith_sif}
K_\mathrm{G} = \sqrt{G_\mathrm{G} E^*}.
\end{equation}

In atomistic simulations of fracture \cite{RefCurtin_review,refErikAtomistic}, a SIF-controlled loading approach is usually employed by displacing atoms according to the linear elastic anisotropic solution in plane strain in mode I \cite{RefLiebowitz}. 
The LEFM displacement field is given by: 
\begin{equation}\label{form_disp_x}
u_{x}(r,\theta) = \frac{K_\mathrm{I}\sqrt{2r}}{\sqrt{\pi}} (f_x(\theta)),
\end{equation}
\begin{equation}\label{form_disp_y}
u_{y}(r,\theta) = \frac{K_\mathrm{I}\sqrt{2r}}{\sqrt{\pi}} (f_y(\theta)),
\end{equation}
where  $K_\mathrm{I}$ is the stress intensity factor under mode I loading and $r$ is the distance of the atom from the mathematical center of the crack tip. The angular distribution functions $f_x(\theta)$ and $f_y(\theta)$ are defined by the angle to the cleavage plane $\theta$ and the elastic constants for a given crystallographic orientation. Although a crack is initially inserted by displacing all atoms, it is only the boundary layers that are kept fixed during the simulations while the remaining atoms are allowed to relax to their minimum energy configuration (see, e.g.,\cite{RefCurtin_review}). 
Furthermore, an explicit failure criterion is not required in atomistic simulations. 
Rather, the fracture toughness $K_{\mathrm{Ic}}$ is a result of these simulations. 
It can be considered to be reached when, as a result of an energy minimization under the applied $K_{\mathrm{Ic}}$ displacement field, 
the separation distance of the crack tip atom pair exceeds some critical value \cite{RefGumbsch,RefJo}.

As the name suggests, a key assumption in LEFM is that the material exhibits linear elastic behavior. Deviations from linear elastic behavior could be due to material nonlinearities, geometrical nonlinearities and time-history dependence \cite{RefClayton}. Material nonlinearities comprise of, for example, nonlinear elastic response or plasticity. Geometrical nonlinearities are due to large deformations where an explicit distinction is to be made between reference and deformed configurations \cite{RefbookSteinmann1}. Deviations due to time-history dependence are usually ascribed to viscoelasticity, creep and fatigue \cite{RefClayton}. 

Previous works with material-specific models such as \cite{RefKermode,RefJo} have shown that there is generally a good agreement between LEFM and atomistics as far as stresses are concerned. The atomic stresses match LEFM away from the crack tip and close to the boundaries where the LEFM displacement field is imposed throughout the simulations. 
Deviations arise only below about 1 nm distance from the crack tip and resolve to finite values. However, these deviations do not invalidate the loading procedure in the simulations since the accompanying changes in energy are localized close to the crack tip \cite{RefCurtin_review}. 
Such deviations are gerenerally ascribed to material nonlinearities, without 
explicitly taking into account possible other nonlinearities.

It should also be noted that the determination of atomic stresses at the crack tip is non-trivial. 
For example, the study by M\"oller et al. \cite{RefJo} observed that the Virial method \cite{refTadmor} used to measure stresses led to deviations at the crack tip due to atomic volumes being ill-defined at surfaces.  

The phenomenon of lattice trapping (which can be generalized to bond trapping for interfaces \cite{refBondTrap}) represents a deviation for atomic structures from Griffith's energy based approach to LEFM (see \cite{refLatticeTrap}). 
The discrete nature of a lattice prevents the continuous increase of crack surfaces by a continuously propagating crack. Instead, cracks propagate by breaking individual atomic bonds. 
This leads to cracks remaining stable above and below the Griffith stress intensity factor $K_\mathrm{G}$ during loading and unloading, respectively. 
While the phenomenon of bond trapping  is well established in the literature (see, e.g, the references in \cite{refErikAtomistic}), the influence of geometrical nonlinearities on lattice trapping have not yet been studied in detail. 

In this study, we provide an example to quantify deviations from the SIF-controlled LEFM displacement field arising from geometric nonlinearities and their contribution to the lattice trapping phenomenon. As a consequence, we show that fracture toughness cannot be analytically determined based on knowledge of maximal bond length of crack tip bonds alone.
Towards this end, we use a harmonic potential with a local cutoff to circumvent nonlinear elastic response prior to cleavage, similar to studies such as \cite{RefGumbsch,RefSinclair}. 
The use of such a potential between atoms is analogous to a "network of springs" model, which is widely used in the study of fracture mechanics \cite{RefPan}.


\section{Method}
\label{method}

The pair force of the harmonic "snapping spring" potential was given by:
\begin{equation}\label{form_harm_force}
F(d) = 
\begin{dcases}
\frac{2 U_{\mathrm{min}}}{[d_\mathrm{0} - d_\mathrm{c}]^2}[d - d_\mathrm{0}], & \text{if } d\leq d_\mathrm{c}\\
0, & \text{otherwise}
\end{dcases}
\end{equation}
where $-U_{\mathrm{min}}$ is the potential energy at the equilibrium distance $d_\mathrm{0}$, $d_\mathrm{c}$ is the cutoff distance, and $d$ is the inter-atomic separation distance. 
This potential leads to a face-centered-cubic (fcc) equilibrium structure.
The parameters and potential properties are listed in table \ref{tab:pot_props}. The cutoff of the potential has to be small enough to be strictly local so that linearity is ensured and that cleavage of bonds at the crack tip can take place \cite{RefSinclair}. Consequently, the critical separation distance between atoms for cleavage coincides with the cutoff $d_\mathrm{c}$. The local nature of the potential also results in the simulations not displaying any surface related phenomena, which are not accounted for in LEFM, making this potential suitable for comparisons with LEFM. Two different parameterization of this harmonic potential have been investigated and resulted in comparable conclusions (the details and results of the second potential are in \ref{pots} and \ref{pot_B}).

\begin{table}[t]
\caption{Summary of parameters and relevant properties of the harmonic potential used in this study (pair potential at equilibrium $U_{\mathrm{min}}$, equilibrium distance $d_\mathrm{0}$, cutoff distance $d_\mathrm{c}$, cohesive energy $E_\mathrm{{coh}}$, lattice constant $a$, surface energy of $(hkl)$ plane $\gamma(hkl)$), elastic constants $C_{ij}$ (cubic symmetry)).}
\label{tab:pot_props}
\centering
\begin{tabularx}{0.45\textwidth}{ 
   >{\centering\arraybackslash}X 
   >{\centering\arraybackslash}X
   >{\centering\arraybackslash}X}
\hline\noalign{\smallskip}
Parameters / Properties & Value \\
\noalign{\smallskip}\hline\noalign{\smallskip}
$U_{\mathrm{min}}$ (eV) & 1.00\\
$d_\mathrm{0}$ (\SI{}{\angstrom}) & 2.54\\
$d_\mathrm{c}$ (\SI{}{\angstrom})& 3.07\\
\\
$E_\mathrm{{coh}}$ (eV) & -6.00\\
$a$ (\SI{}{\angstrom}) & 3.60\\
$\gamma(100)$ (J/m\textsuperscript{2}) & 4.95\\
$C_{11}$ (GPa) & 642.92\\
$C_{12}$ (GPa) & 321.59\\
$C_{44}$ (GPa) & 321.59\\
\noalign{\smallskip}\hline
\end{tabularx}
\end{table}

\begin{figure}[h]
\centering
  \includegraphics[width=0.45\textwidth]{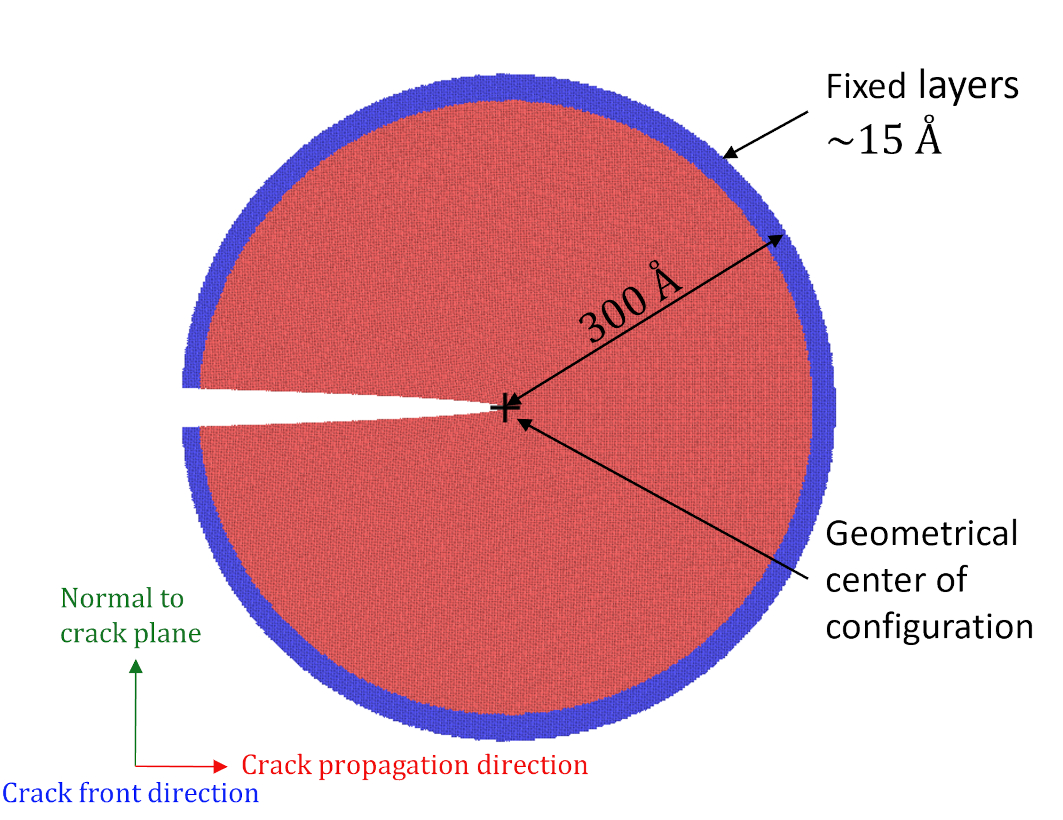}
\caption{Schematic of the simulation setups used. In this study, the geometrical center of the configurations coincided with the mathematical center of the LEFM displacement field in all simulations. Red atoms are free to move, blue atoms are fixed.}
\label{fig:setup}
\end{figure}

Cylindrical pacman-like configurations with radii of \SI{300}{\angstrom} and depths of about \SI{10}{\angstrom} were used, see Fig. \ref{fig:setup}. Without removing atoms or deleting bonds, LEFM near crack tip solutions according to equations (\ref{form_disp_x}) and (\ref{form_disp_y}) were used to create and load cracks. A boundary region of about \SI{15}{\angstrom} thickness was kept fixed throughout the simulations, whereas the remaining domain was allowed to relax. The sizes of the configurations were large enough to converge to the infinite limit assumed by LEFM with respect to $K_{\mathrm{Ic}}$ (see \ref{tests}). Hence, flexible boundary conditions as detailed in \cite{RefKermode_arcLength} were not required. Instead, the LEFM displacement field and boundaries were always centered on the geometrical center of the configurations, similar to works such as \cite{RefJo, RefCurtin, RefLars}. The crack systems (characterized by crack plane and crack front direction) studied were (100)[001] and (100)[011].

The initial prescriptions of the displacement field were done at a load where the cracks were stable at the center of the configurations ($K_{\mathrm{in}}$), see
table \ref{tab:k-values} for the values of $K_G$ and $K_{\mathrm{in}}$.
The configurations were then minimized using FIRE \cite{RefFireLammps}. Two procedures were used in this study to impose further loads: {\em incremental} and {\em total} loading. 
In incremental loading, further loads $\Delta K$ in the form of prescribed displacements $\Delta \boldsymbol{u}$ according to equations (\ref{form_disp_x}) and (\ref{form_disp_y}) were incrementally applied on all the atoms in the relaxed configuration of the previous load increment until the crack tip bonds underwent cleavage ($d> d_\mathrm{c}$). The fracture toughness determined that way are referred to by $K_{\mathrm{Ic}}^\mathrm{inc}$. 
In the case of total loading, displacements $\boldsymbol{u}$ according to the total desired load $K_{\mathrm{in}}+\Delta K$ were directly applied to the initial, uncracked cylinder.
The sample were then relaxed and the procedure was repeated, until the separation distance of the crack tip atom pair exceeded $d_\mathrm{c}$. The fracture toughness determined by total loading is $K_{\mathrm{Ic}}^\mathrm{tot}$.

All calculations were performed with LAMMPS \cite{RefLAMMPS} and analysis was done with the help of OVITO \cite{RefOVITO}.


\section{Results and Discussion}
\label{rnd}

\begin{figure}[h]
\centering
  \includegraphics[width=0.45\textwidth]{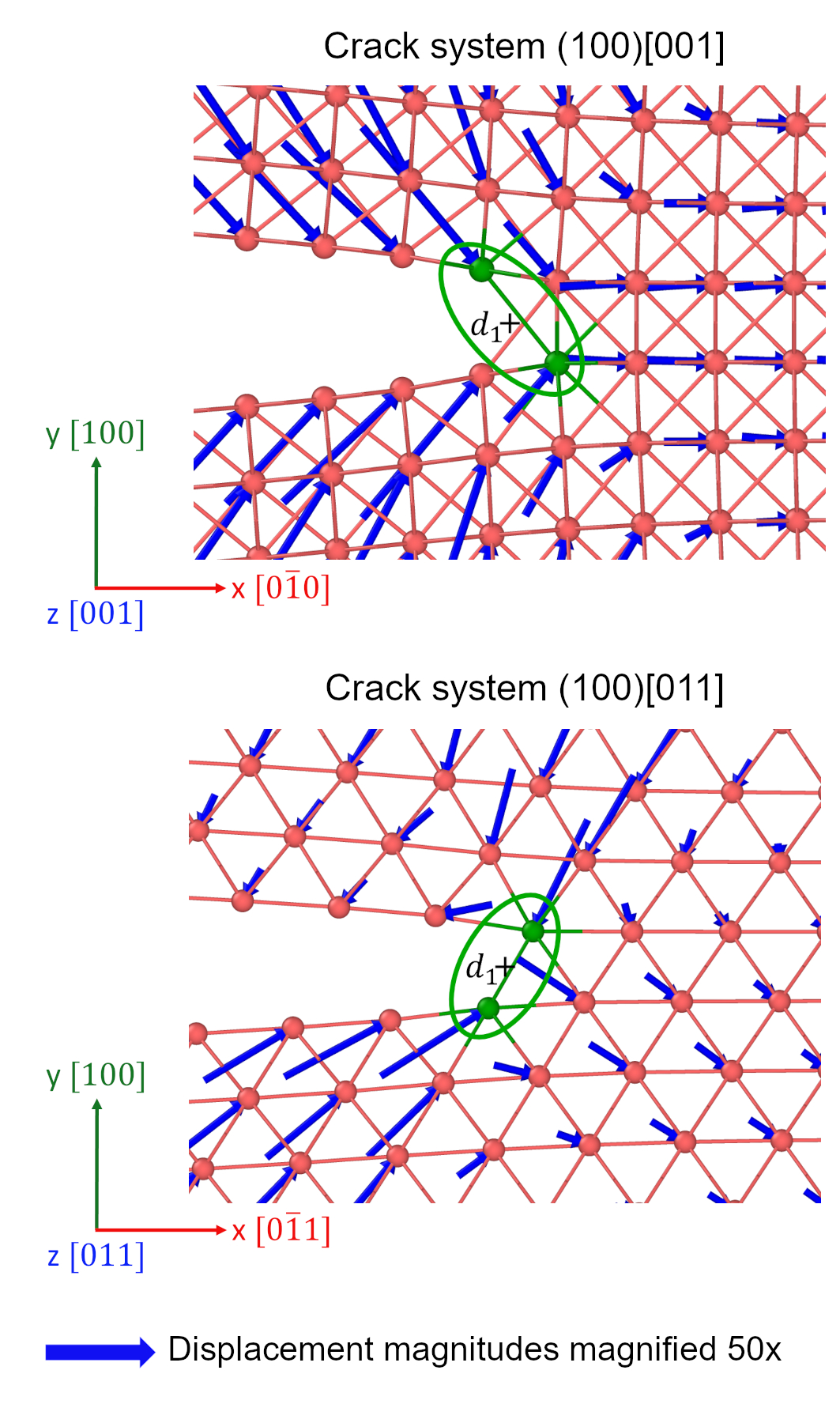}
\caption{Positions of atoms around crack tips after relaxation at $K_{\mathrm{in}}$. The blue arrows show the (magnified) difference between the atom positions according to LEFM (Eqs. (\ref{form_disp_x}) and (\ref{form_disp_y}) and after relaxation. Similar deviating displacements are observed at all higher loads with both loading procedures. The crack tip atom pair is circled and highlighted by green atoms.}
\label{fig:disp}
\end{figure}

Prior to cleavage of crack tip bonds, both crack systems with both loading procedures show deviations from the LEFM prescribed positions due to relaxation, see Fig. \ref{fig:disp}. These deviating displacements indicate inconsistency with the linear behavior assumed by LEFM. Since the elastic response of the material before cleavage is linear, these deviations can only be ascribed to geometrical nonlinearities. In other words, collective relaxation processes ahead of the crack tip lead to changes in geometry of the crack tip neighborhood which is no longer compatible with linear elasticity.

\begin{figure}[h]
\centering
  \includegraphics[width=0.45\textwidth]{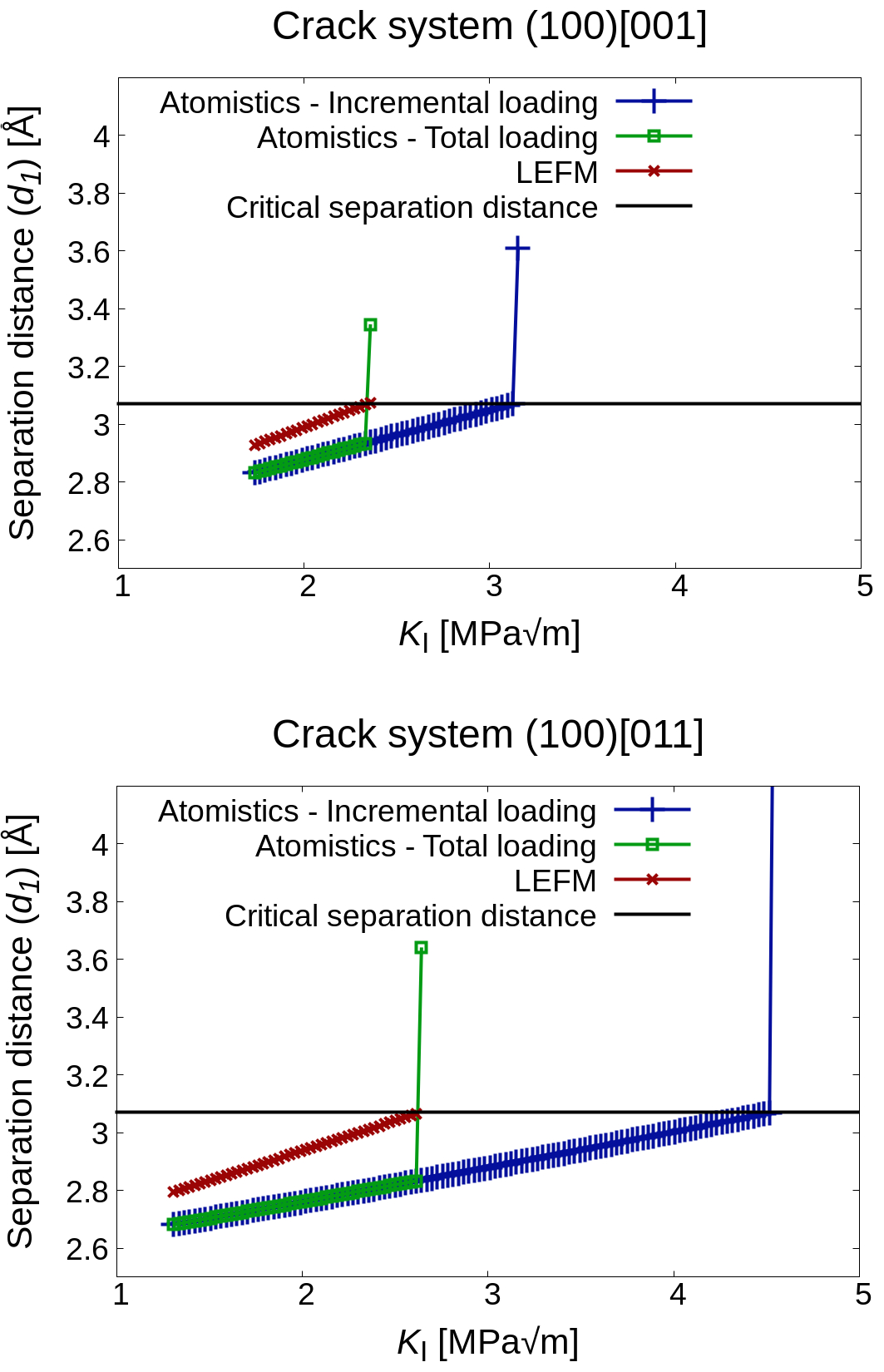}
\caption{Separation distance of crack tip atoms (see Fig.\ref{fig:disp}) as a function of $K_\mathrm{I}$ according to LEFM and the two loading procedures.}
\label{fig:sepdist}
\end{figure}

Although these deviating displacements seem minor, they have a significant impact on the separation distances of the crack tip atom pairs. It can be seen from Fig. \ref{fig:sepdist} that large deviations exist in both crack systems between the separation distance according to the atomic positions determined by LEFM and the atomistic response, independent of loading procedure. 
In the case of total loading, the crack tip bonds, however, cleave in accordance with the analytical LEFM equations (\ref{form_disp_x}) and (\ref{form_disp_y}). 
This is due to the atoms being positioned by the LEFM displacement field so that the crack tip bonds are already cleaved at $K_{\mathrm{Ic}}^\mathrm{tot}$, and the bonds do not heal during minimization.
Until then, the separation distance $d_1$ remains nearly identical for both loading procedures. 
In the case of incremental loading, the deviations add up and lead to fracture toughness values that are $>30\%$ larger than the corresponding analytically determined LEFM values.

The material shows high lattice trapping as evidenced by $K_{\mathrm{Ic}}^\mathrm{inc} >> K_\mathrm{G}$.
This is a consequence of its local and linear nature \cite{RefSinclair, RefGumbsch}. This inherent increase in the critical SIF is represented in the total loading procedure. From table \ref{tab:k-values}, it can be seen that critical SIFs with total loading $K_{\mathrm{Ic}}^\mathrm{tot}$ are substantially larger ($>70\%$) than values based on the Griffith criterion $K_\mathrm{G}$. However, the critical SIFs with incremental loading $K_{\mathrm{Ic}}^\mathrm{inc}$ are even larger due to the deviating displacements from relaxation. Thus, for incremental loading, geometric nonlinearities significantly influence lattice trapping.

\begin{table}[!b]
\caption{Comparisons of initial load ($K_\mathrm{in}$), critical stress intensity factor from Griffith's criterion ($K_\mathrm{G}$), critical stress intensity factors from incremental loading ($K_{\mathrm{Ic}}^\mathrm{inc}$), critical stress intensity factors from and total loading ($K_{\mathrm{Ic}}^\mathrm{tot}$). All values are in $\mathrm{MPa}\sqrt{\mathrm{m}}$.}
\label{tab:k-values}
\centering
\begin{tabularx}{0.45\textwidth}{ 
   >{\centering\arraybackslash}X 
   >{\centering\arraybackslash}X
   >{\centering\arraybackslash}X}
\hline\noalign{\smallskip}
 & (100)[001] & (100)[011] \\
\noalign{\smallskip}\hline\noalign{\smallskip}
$K_\mathrm{in}$ & 1.70 & 1.28\\
\\
$K_\mathrm{G}$ & 1.39 & 1.42\\
\\
$K_{\mathrm{Ic}}^\mathrm{inc}$ & 3.15 & 4.54\\
\\
$K_{\mathrm{Ic}}^\mathrm{tot}$ & 2.36 & 2.64\\
\noalign{\smallskip}\hline
\end{tabularx}
\end{table}

However, material-specific, more realistic material models show much lower lattice trapping,
see, e.g., the work by Hiremath et al. \cite{refOlsson} on cracks in Tungsten under incremental loading. 
There it was found with a newly developed, DFT fitted, modified embedded atom method potential
that the value of $K_{\mathrm{Ic}}^\mathrm{inc}$ was just $2\%$ larger than $K_\mathrm{G}$ ((001)[1-10] crack system).  

They also published the traction-separation curves.
The position of the peak of these curves ($\delta$) can be considered for the critical separation distance ($d_\mathrm{c}=d_{(010)}+\delta$) for a vertically orientated crack tip bond (as is the case in this orientation).
One can then use the  LEFM equations (\ref{form_disp_x}) and (\ref{form_disp_y})
to determine $K_\mathrm{Ic}(d=d_\mathrm{c})$. 
With $\delta\approx \SI{0.5}{\angstrom}$ and the values of $d_{(010)}$ and the elastic constant 
for the potential, see \cite{refOlsson}, the so calculated $K_\mathrm{Ic}$ is however about 125\% larger than the measured one. 
This again highlights that using only the critical bond separation distance is 
not sufficient to calculate the fracture toughness with LEFM.
In this example, however, not only geometrical nonlinearities are at play,
but also material nonlinearities and surface effects like surface relaxation.

\begin{figure}[!h]
\centering
  \includegraphics[width=0.45\textwidth]{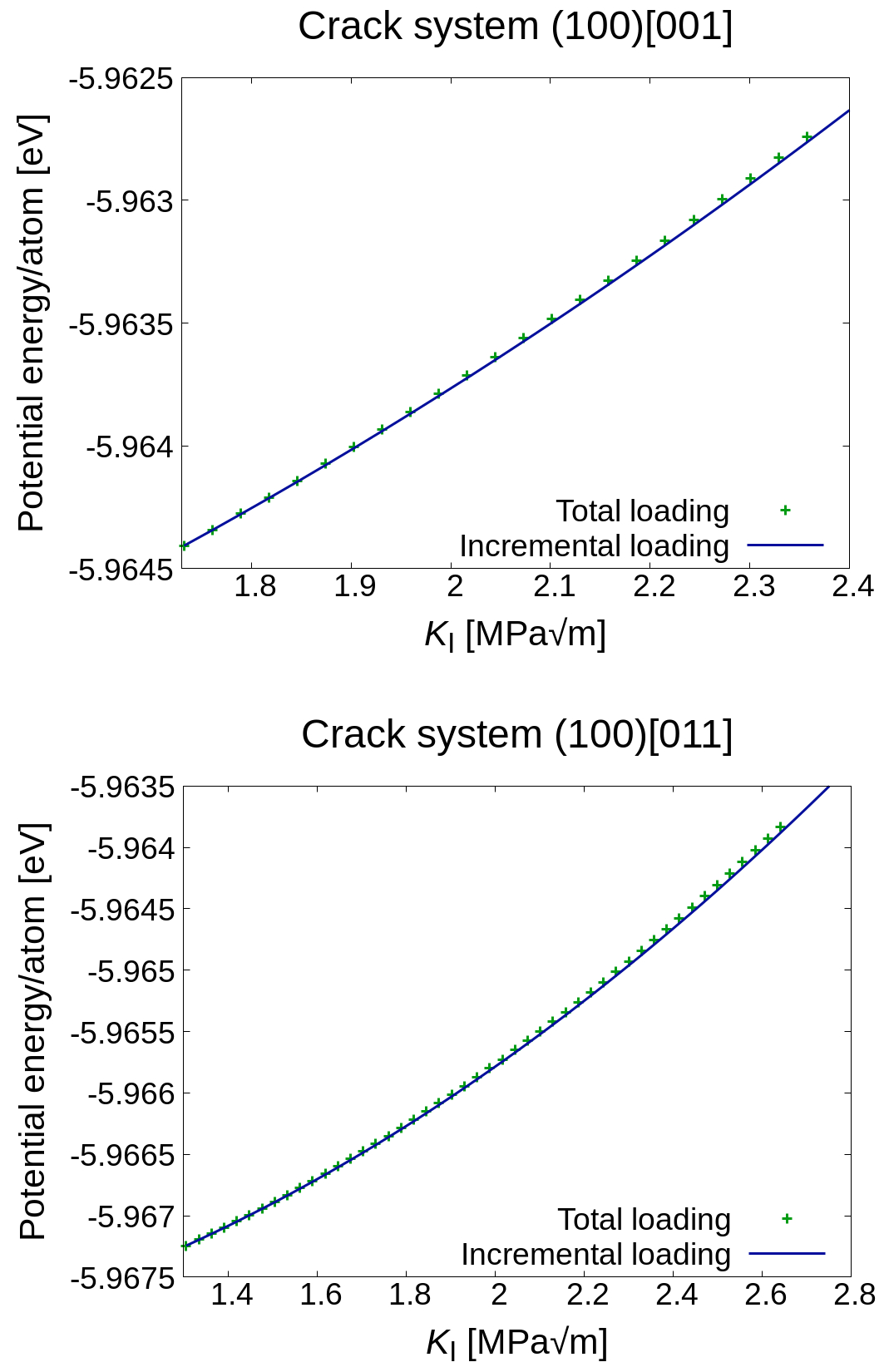}
\caption{Potential energy per atom as a function of $K_\mathrm{I}$. The plots are focused around the loads at which the crack tip bonds underwent cleavage with the total loading procedure ($K_{\mathrm{Ic}}^\mathrm{tot}$).}
\label{fig:eng}
\end{figure}

Finally, the two loading procedures are compared. As noted by Sinclair \cite{RefSinclair}, the transition from a pristine crystal to a fractured surface will not be sudden. Hence, the fracture toughness values from total loading may not be realistic (when used with realistic materials). 
The two loading procedures start at the same $K_\mathrm{in}$, at which
the corresponding structures have identical total energies.
As shown in Fig. \ref{fig:eng}, with further loading the energies 
deviate from each other, however, at $K_\mathrm{Ic}^\mathrm{tot}$ the 
difference is less than 1 meV per atom in both crack systems.
This energy difference seems relatively low, however, the 
structural difference are located close to the crack tip and can therefore
play an important role.
 If fracture is assumed to be sufficiently slow so that atoms have time to find their minimum energy configuration, the procedure that provides the lowest energies for the given load $K_\mathrm{I}$ would have to be considered.

\section{Conclusions}
\label{conclusions}

The results presented in this work show that geometrical nonlinearities cause deviations from LEFM even if the material has a linear elastic response prior to bond cleavage. 
In the case of incremental loading, which can be assumed to be relevant in the determination of crack initiation toughness, geometrical nonlinearities also influence lattice trapping.
Real materials further deviate from LEFM due to their nonlinear elastic responses and surface effects. 
We show, however, that even in the absence of such complications, fracture toughness cannot be re-conciliated with LEFM by using the critical separation distance of the crack tip atom pair in an incremental loading procedure.  
Rather, fracture is a collective phenomenon at the atomic scale, even with purely local and linear interactions.
Therefore actual fracture simulations have to be performed to determine fracture toughness.


\appendix

\section{Potentials}
\label{pots}

The formulation of the potentials is given by
\begin{equation}\label{form_harm_pot1}
V(d) = \frac{U_{\mathrm{min}}}{[d_\mathrm{0} - d_\mathrm{c}]^2} \left[ [d - d_\mathrm{0}]^2 - [d_\mathrm{c} - d_\mathrm{0}]^2 \right],
\end{equation}
where, $V$ is the effective pair potential, with $-U_{min}$ being the potential at the equilibrium distance $d_0$, $d_c$ being the cutoff distance, and $d$ being the inter-atomic separation distance. The pair force is then given by the first derivative of (\ref{form_harm_pot1}):
\begin{equation}\label{form_harm_force1}
F(d) = \frac{2 U_{\mathrm{min}}}{[d_\mathrm{0} - d_\mathrm{c}]^2}[d - d_\mathrm{0}].
\end{equation}

Two potentials labelled "Harmonic-A" and "Harmonic-B" were used, with the properties listed in table \ref{tab:pot_props}. The results of Harmonic-A are presented in the main manuscript. The results of "Harmonic-B" are in section \ref{pot_B}.

\begin{table}[h]
\caption{Summary of parameters and relevant properties of the harmonic potentials (pair potential at equilibrium $U_{\mathrm{min}}$, equilibrium distance $d_\mathrm{0}$, cutoff distance $d_\mathrm{c}$, cohesive energy $E_\mathrm{{coh}}$, lattice constant $a$, surface energy of $(hkl)$ plane $\gamma(hkl)$), elastic constants $C_{ij}$ (cubic symmetry)).}
\label{tab:pot_props_App}
\begin{tabularx}{0.45\textwidth}{ 
   >{\centering\arraybackslash}X 
   >{\centering\arraybackslash}X
   >{\centering\arraybackslash}X}
\hline\noalign{\smallskip}
Parameters / Properties & Harmonic-A & Harmonic-B \\
\noalign{\smallskip}\hline\noalign{\smallskip}
$U_\mathrm{{min}}$ (eV) & 1.00 & 1.00\\
$d_\mathrm{0}$ (\SI{}{\angstrom}) & 2.54 & 2.54\\
$d_\mathrm{c}$ (\SI{}{\angstrom})& 3.07 & 2.90\\
\\
$E_\mathrm{{coh}}$ (eV) & -6.00 & -6.00\\
$a$ (\SI{}{\angstrom}) & 3.597 & 3.597\\
$\gamma(100)$ (J/m\textsuperscript{2}) & 4.95 & 4.95\\
$C_{11}$ (GPa) & 642.92 & 1401.76\\
$C_{12}$ (GPa) & 321.59 & 700.95\\
$C_{44}$ (GPa) & 321.59 & 700.79\\
\noalign{\smallskip}\hline
\end{tabularx}
\end{table}


\section{Tests of setup and simulation parameters}
\label{tests}

Convergence of $K_{\mathrm{Ic}}^\mathrm{inc}$ with respect to configuration radius $R$ was tested with Harmonic-A using the incremental loading procedure (see table \ref{tab:config_sizes}). It can be seen that $R=\SI{300}{\angstrom}$ was sufficient for both crack systems, with further increase in configuration size resulting in negligible change to $K_{\mathrm{Ic}}$. Simulations of Harmonic-B were done only with $R=\SI{300}{\angstrom}$.

\begin{table}[!h]
\caption{$K_{\mathrm{Ic}}^\mathrm{inc}$ ($\mathrm{MPa}\sqrt{\mathrm{m}}$) values of crack systems with varying configuration radii ($R$) using Harmonic-A. The convergence threshold (fnorm-thr) was 1e-6 eV/\SI{}{\angstrom} and load increment ($\Delta K_\mathrm{I}$) was $0.028 \ \mathrm{MPa}\sqrt{\mathrm{m}}$.}
\label{tab:config_sizes}
\begin{tabularx}{0.45\textwidth}{ 
   >{\centering\arraybackslash}X 
   >{\centering\arraybackslash}X
   >{\centering\arraybackslash}X
   >{\centering\arraybackslash}X}
\hline\noalign{\smallskip}
Crack system & $R=\SI{150}{\angstrom}$ & $R=\SI{300}{\angstrom}$ & $R=\SI{600}{\angstrom}$ \\
\noalign{\smallskip}\hline\noalign{\smallskip}
(100)[001] & 3.12 & 3.15 & -\\
(100)[011] & 4.48 & 4.54 & 4.57\\
\noalign{\smallskip}\hline
\end{tabularx}
\end{table}

Influence of convergence threshold (fnorm-thr) was tested with Harmonic-A using the incremental loading procedure (see table \ref{tab:fnorm_thr}). It can be seen that a convergence threshold of fnorm-thr = 1e-6 eV/\SI{}{\angstrom} was sufficient, and using a tighter threshold produced no change in $K_{\mathrm{Ic}}^\mathrm{inc}$. Simulations with Harmonic-B were performed only with fnorm-thr = 1e-6 eV/\SI{}{\angstrom}. 

\begin{table}[!h]
\caption{$K_{\mathrm{Ic}}^\mathrm{inc}$ ($\mathrm{MPa}\sqrt{\mathrm{m}}$) values of crack systems for varying convergence thresholds (fnorm-thr) with Harmonic-A. The configuration radius ($R$) was \SI{300}{\angstrom} and load increment ($\Delta K_\mathrm{I}$) was $0.028 \ \mathrm{MPa}\sqrt{\mathrm{m}}$. }
\label{tab:fnorm_thr}
\begin{tabularx}{0.45\textwidth}{ 
   >{\centering\arraybackslash}X 
   >{\centering\arraybackslash}X
   >{\centering\arraybackslash}X}
\hline\noalign{\smallskip}
Crack system & fnorm-thr = 1e-6 eV/\SI{}{\angstrom} & fnorm-thr = 1e-8 eV/\SI{}{\angstrom}\\
\noalign{\smallskip}\hline\noalign{\smallskip}
(100)[001] & 3.15 & 3.15\\
(100)[011] & 4.54 & 4.54\\
\noalign{\smallskip}\hline
\end{tabularx}
\end{table}

Influence of loading increment ($\Delta K_\mathrm{I}$) on $K_{\mathrm{Ic}}^\mathrm{inc}$ ($\mathrm{MPa}\sqrt{\mathrm{m}}$) was tested with Harmonic-A using the incremental loading procedure (see table \ref{tab:kinc}). It can be seen that having smaller increments than $0.028 \ \mathrm{MPa}\sqrt{\mathrm{m}}$ has little influence, whereas it increases computational cost (more steps needed). Hence, $\Delta K_\mathrm{I} = 0.028 \ \mathrm{MPa}\sqrt{\mathrm{m}}$ was used (also for Harmonic-B).

\begin{table}[!h]
\caption{$K_{\mathrm{Ic}}^\mathrm{inc}$ ($\mathrm{MPa}\sqrt{\mathrm{m}}$) values of crack systems for varying load increments ($\Delta K_\mathrm{I}$) with Harmonic-A. The configuration radius ($R$) was \SI{300}{\angstrom} and the convergence threshold (fnorm-thr) was 1e-6 eV/\SI{}{\angstrom}.}
\label{tab:kinc}
\begin{tabularx}{0.45\textwidth}{ 
   >{\centering\arraybackslash}X 
   >{\centering\arraybackslash}X
   >{\centering\arraybackslash}X}
\hline\noalign{\smallskip}
Crack system & $\Delta K_\mathrm{I} = 0.028 \ \mathrm{MPa}\sqrt{\mathrm{m}}$ & $\Delta K_\mathrm{I} = 0.014 \ \mathrm{MPa}\sqrt{\mathrm{m}}$\\
\noalign{\smallskip}\hline\noalign{\smallskip}
(100)[001] & 3.15 & 3.14\\
(100)[011] & 4.54 & 4.53\\
\noalign{\smallskip}\hline
\end{tabularx}
\end{table}


\section{Results with Harmonic-B}
\label{pot_B}

\begin{figure}[!hb]
  \includegraphics[width=0.45\textwidth]{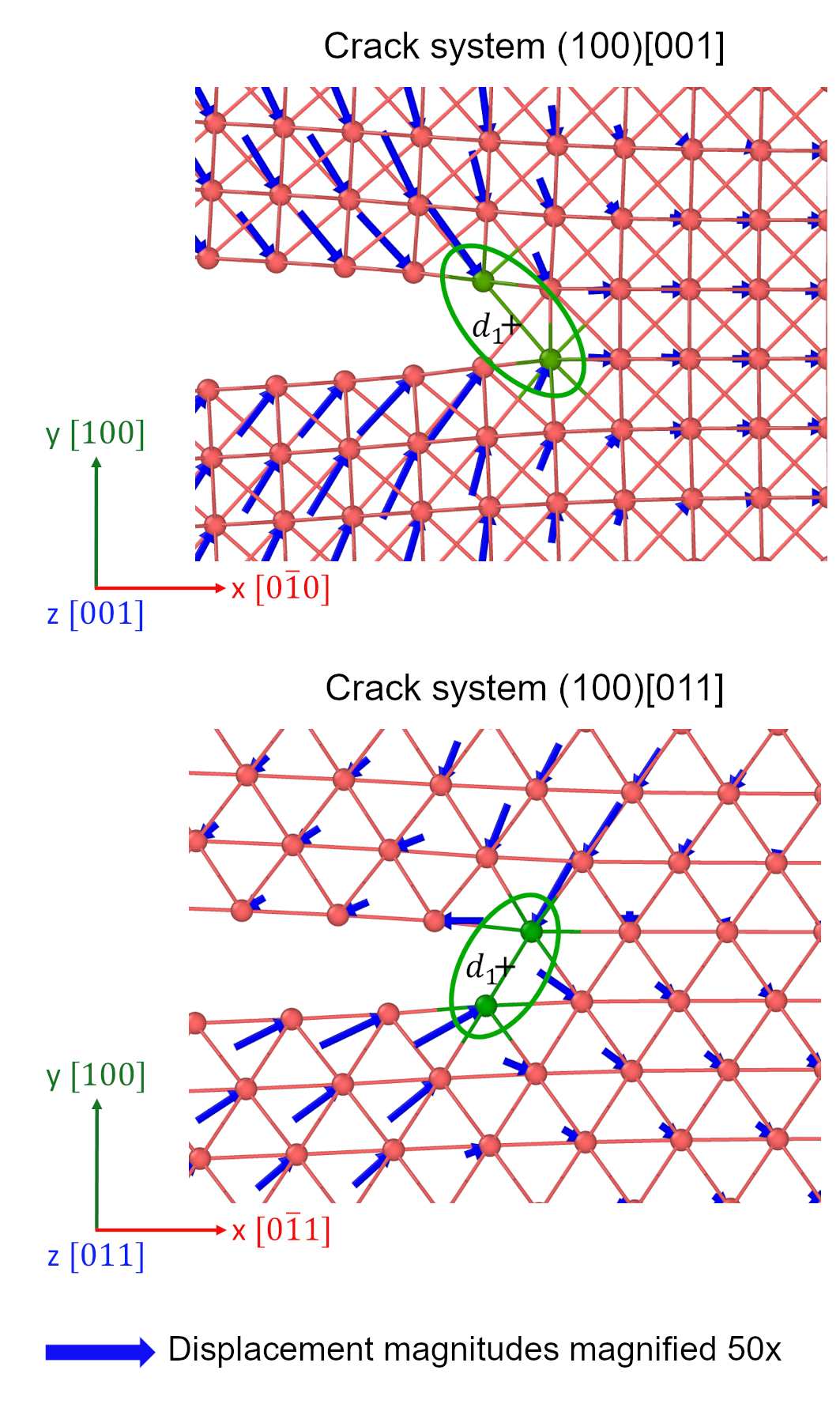}
\caption{Positions of atoms around crack tips after relaxation at $K_{\mathrm{in}}$ with Harmonic-B. The blue arrows show the (magnified) difference between the atom positions according to LEFM (Eqs. (\ref{form_disp_x}) and (\ref{form_disp_y}) and after relaxation. Similar deviating displacements are observed at all higher loads with both loading procedures. The crack tip atom pair is circled and highlighted by green atoms.}
\label{fig:displ_harmonicB}
\end{figure}

\begin{figure}[!b]
  \includegraphics[width=0.45\textwidth]{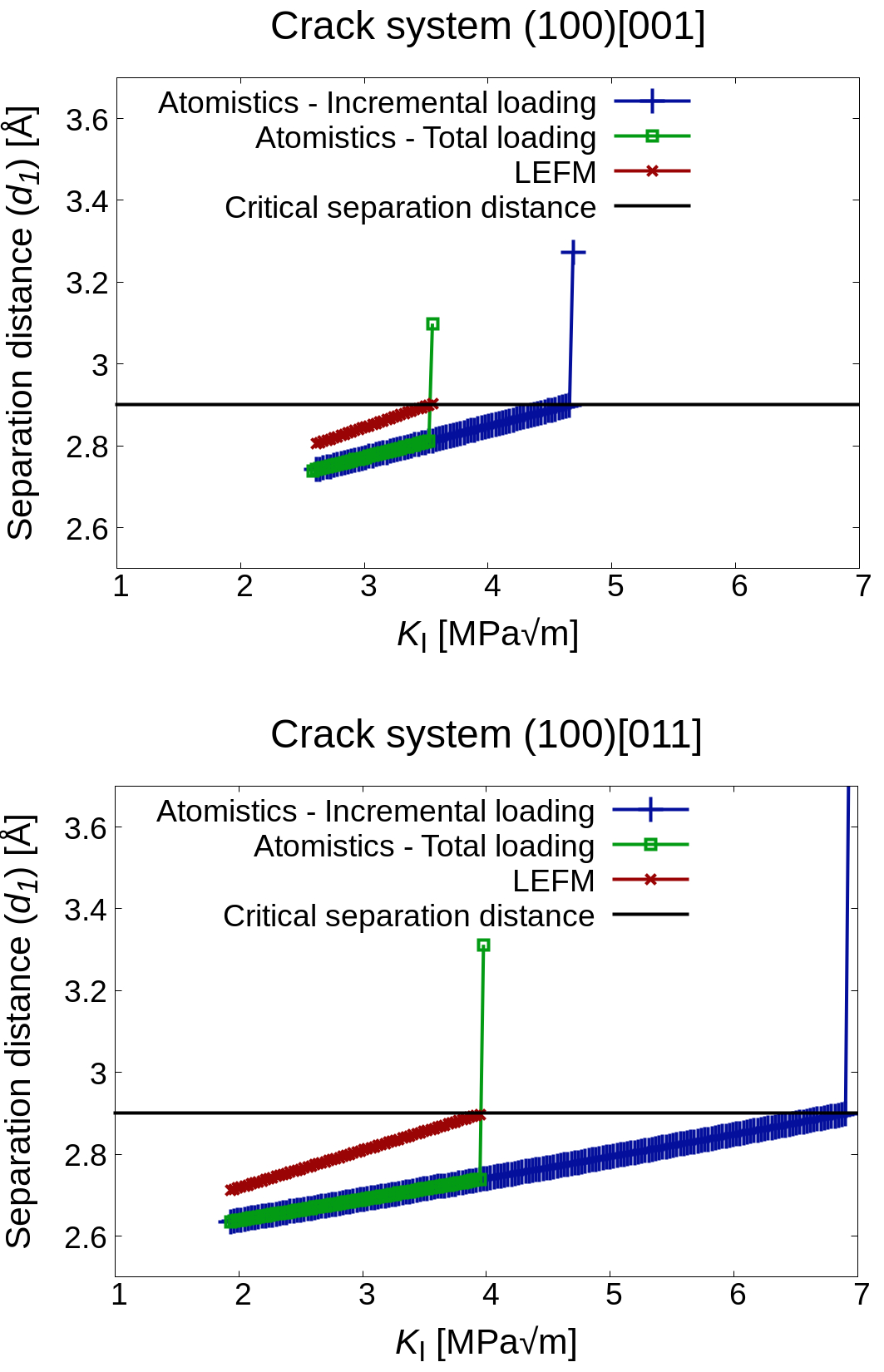}
\caption{Separation distance of crack tip atoms (see Fig.\ref{fig:displ_harmonicB}) with Harmonic-B as a function of $K_\mathrm{I}$ according to LEFM and the two loading procedures.}
\label{fig:sepdist_harmonicB}
\end{figure}

\begin{figure}[h]
  \includegraphics[width=0.45\textwidth]{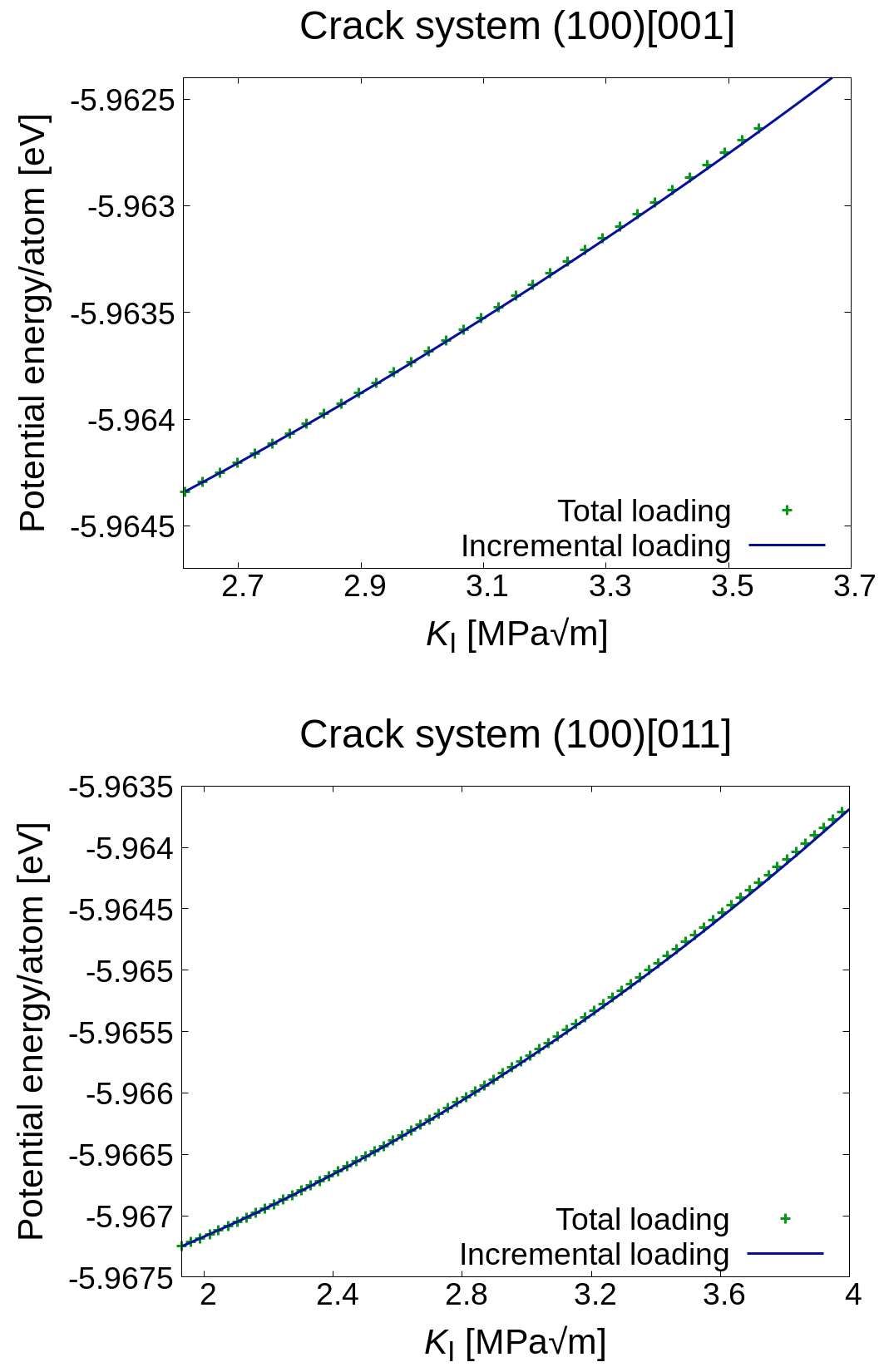}
\caption{Potential energy per atom as a function of $K_\mathrm{I}$ with Harmonic-B. The plots are focused around the loads at which the crack tip bonds underwent cleavage with the total loading procedure.}
\label{fig:energiesB} 
\end{figure}

The results of the simulations with Harmonic-B are qualitatively similar to Harmonic-A with respect to displacements from LEFM-prescribed positions during relaxations, as well as with respect to fracture toughness values and energetics with both loading procedures (see fig.\ref{fig:displ_harmonicB}, fig.\ref{fig:sepdist_harmonicB} and fig. \ref{fig:energiesB}).

\section*{Acknowledgements}
\label{acknowledgements}
This research was funded by the Deutsche Forschungsgemeinschaft (DFG, German Research Foundation) - 377472739/GRK 2423/1-2019. EB further acknowledges support from the European Research Council (ERC) under the European Union’s Horizon 2020 research and innovation programme (grant agreement No 725483).

\bibliographystyle{elsarticle-num} 
\bibliography{2021_Paper_Fracture_abInitio_harmonic}





\end{document}